\documentclass[twocolumn,showpacs,preprintnumbers,superscriptaddress,amsmath,amssymb,pre]{revtex4}

\usepackage{graphicx}
\usepackage{dcolumn}
\usepackage{bm}
\usepackage{color}

\begin{document}



\title{Modified detrended fluctuation analysis based on empirical mode decomposition}

\author{Xi-Yuan Qian}
 \email{xyqian@ecust.edu.cn}
 \affiliation{School of Science, East China University of Science and Technology, Shanghai 200237, China} %
 \affiliation{Research Center for Econophysics, East China University of Science and Technology, Shanghai 200237, China} %

\author{Wei-Xing Zhou}
 \email{wxzhou@ecust.edu.cn}
 \affiliation{School of Science, East China University of Science and Technology, Shanghai 200237, China} %
 \affiliation{Research Center for Econophysics, East China University of Science and Technology, Shanghai 200237, China} %
 \affiliation{School of Business, East China University of Science and Technology, Shanghai 200237, China} %
 \affiliation{Engineering Research Center of Process Systems Engineering (Ministry of Education), East China University of Science and Technology, Shanghai 200237, China}
 \affiliation{Research Center on Fictitious Economics \& Data Science, Chinese Academy of Sciences, Beijing 100080, China}

\author{Gao-Feng Gu}
 \email{gfgu@ecust.edu.cn}
 \affiliation{School of Science, East China University of Science and Technology, Shanghai 200237, China} %
 \affiliation{Research Center for Econophysics, East China University of Science and Technology, Shanghai 200237, China} %
 \affiliation{School of Business, East China University of Science and Technology, Shanghai 200237, China} %

\date{\today}

\begin{abstract}
Detrended fluctuation analysis (DFA) is a simple but very efficient
method for investigating the power-law long-term correlations of
non-stationary time series, in which a detrending step is necessary
to obtain the local fluctuations at different timescales. We propose
to determine the local trends through empirical mode decomposition
(EMD) and perform the detrending operation by removing the EMD-based
local trends, which gives an EMD-based DFA method. Similarly, we
also propose a modified multifractal DFA algorithm, called an
EMD-based MFDFA. The performance of the EMD-based DFA and MFDFA
methods is assessed with extensive numerical experiments based on
fractional Brownian motion and multiplicative cascading process. We
find that the EMD-based DFA method performs better than the classic
DFA method in the determination of the Hurst index when the time
series is strongly anticorrelated and the EMD-based MFDFA method
outperforms the traditional MFDFA method when the moment order $q$
of the detrended fluctuations is positive. We apply the EMD-based
MFDFA to the one-minute data of Shanghai Stock Exchange Composite
index, and the presence of multifractality is confirmed.
\end{abstract}

\pacs{05.45.Tp, 05.40.-a, 05.45.Df, 89.65.Gh}

\maketitle



\section{Introduction}
\label{s1:Introduction}

The dynamics of an evolving complex system can be usually recorded
as time series, whose temporal correlation structure embeds much
information about the interactions among the microscopic
constituents of the system. There are a wealth of approaches
proposed to determine the correlation strength of a time series
\cite{Taqqu-Teverovsky-Willinger-1995-Fractals,Montanari-Taqqu-Teverovsky-1999-MCM}.
In recent years, the detrended fluctuation analysis (DFA) has become
the most extensively adopted method. The idea of DFA was originally
invented to investigate the long-range dependence in coding and
noncoding DNA nucleotide sequences
\cite{Peng-Buldyrev-Havlin-Simons-Stanley-Goldberger-1994-PRE,Kantelhardt-Bunde-Rego-Havlin-Bunde-2001-PA}.
An extension of the DFA method can be used to unveil the
multifractal nature hidden in time series, termed the multifractal
DFA (MFDFA)
\cite{Kantelhardt-Zschiegner-Bunde-Havlin-Bunde-Stanley-2002-PA}.
The (MF)DFA method can also be generalized to investigate
higher-dimensional fractal and multifractal measures
\cite{Gu-Zhou-2006-PRE}.

There are also other applications of the DFA method. One example is
the investigation of lagged correlations of nonstationary signals,
in which the so-called {\em{lagged}} DFA can determine the largest
correlation uncovering the existence of underlying delays in the
evolution of real time series
\cite{AlvarezRamirez-Rodriguez-Echeverria-2009-PRE}. On the other
hand, there are many situations where several variables are
simultaneously recorded that exhibit long-range dependence or
multifractal nature. A detrended cross-correlation analysis (DXA)
was proposed to investigate the long-range cross-correlations
between two nonstationary time series, which is a generalization of
the DFA method \cite{Podobnik-Stanley-2008-PRL,Zhou-2008-PRE}. These
methods can be easily generalized to study the power-law
cross-correlations in higher-dimensional fractal or multifractal
signals.

A common step of all the aforementioned methods is to remove the
local trends at different timescales. However, for a real time
series, it is usually not known if there is a trend and, if any,
what the functional form of this trend is. A conventional strategy
is to assume that the local trends are in the form of polynomial,
which works quite well. Recently, a model-free timescale-adaptive
detrending approach has been proposed, which is based on the
empirical mode decomposition (EMD) approach
\cite{Wu-Huang-Long-Peng-2007-PNAS}. For a given time series, EMD is
designed to decompose it into a limited number of intrinsic mode
functions (IMFs) and a residue component. The EMD-based trend can be
considered as the combination of the monotonic residue and IMFs
which are significantly distinguishable from the pure white noise,
and the timescale of an EMD-based trend is the averaged timescale of
the decomposed IMFs. In this work, we employ the EMD as a detrending
tool to modify the DFA algorithm, as well as the MFDFA algorithm.

There are several relevant studies aiming at applying the EMD to
study the correlations in time series. Numerical experiments based
on fractional Gaussian noise with Hurst index $H_0$ show that the
EMD acts as a dyadic filter bank which is able to extract the Hurst
index $H$ in good agreement with the ``true value'' $H_0$ when
$H_0\geqslant0.5$ but shows a clear deviation from $H_0$ when
$H_0<0.5$ \cite{Flandrin-Rilling-Goncalves-2004-IEEspl}. A close
scrutiny unveils that the extracted $H$ is systematically greater
than $H_0$ \cite{Flandrin-Rilling-Goncalves-2004-IEEspl}. Another
idea is to perform EMD on the original time series to remove the
trend or seasonality and a DFA is followed, which was applied to
study the correlation properties of long daily ozone records
\cite{Janosi-Muller-2005-PRE}. A slight different method is to
conduct the EMD detrending process and the DFA directly on the
original time series and a two-parameter scale of randomness for the
DFA is proposed to replace the DFA scaling exponent, which was
applied to identify characteristics and complexity of heartbeat
interval caused by the effects of aging and of illness
\cite{Yeh-Fan-Shieh-2009-MEP}. Indeed, there is no power-law
dependence of the detrended fluctuation function on the timescale.

As we will show below, our EMD-based DFA and MFDFA algorithms are
different from the previous efforts in the combination of EMD and
DFA. The performance of the EMD-based DFA and MFDFA methods is
assessed with numerical experiments based on fractional Brownian
motion and multiplicative cascading process. We find that the
EMD-based DFA method performs better than the classic DFA method in
the determination of the Hurst index when the time series is
strongly anticorrelated, and the EMD-based MFDFA method outperforms
the traditional MFDFA method when the moment order $q$ of the
detrended fluctuations is positive.

The paper is organized as follows. In Sec.~\ref{s1:DFAandEMD}, we
review the algorithms of DFA, MFDFA and EMD and propose the
EMD-based DFA and MFDFA algorithms. In Sec.~\ref{s1:Simulation}, we
test the performance of the EMD-based DFA and MFDFA algorithms with
extensive numerical experiments. The EMD-based MFDFA method is
applied in Sec.~\ref{S1:Application} to investigate the multifractal
nature of the return time series of the Shanghai Stock Exchange
Composite index. Section~\ref{s1:Conclusion} gives a brief summary.

\section{The EMD-based DFA and MFDFA algorithms}
\label{s1:DFAandEMD}

\subsection{The DFA algorithm}

The original DFA algorithm contains the following five steps
\cite{Peng-Buldyrev-Havlin-Simons-Stanley-Goldberger-1994-PRE,Kantelhardt-Bunde-Rego-Havlin-Bunde-2001-PA}.

{\em{Step 1}}. Consider a time series $x(t)$, $t=1,2,\cdots,N$.
First construct the cumulative sum
\begin{equation}
u(t) = \sum_{i = 1}^{t}{x(i)}, ~~t=1,2,\cdots,N~.
 \label{Eq:dfa_u}
\end{equation}

{\em{Step 2}}. The new series $u(t)$ is partitioned into $N_s$
disjoint segments of the same size $s$, where $N_s = [N/s]$. Each
segment can be denoted by $u_v$ such that $u_v(i)=u(l+i)$ for
$1\leqslant{i}\leqslant{s}$, where $l=(v-1)s$.

{\em{Step 3}}. In each segment $u_v$, we determine the local trend
$\widetilde{u}_v$ with the method of polynomial fitting. When a
polynomial of order $\ell$ is adopted in this step, the DFA method
is called DFA-$\ell$ (DFA-1 if $\ell$=1, DFA-2 if $\ell$=2, DFA-3 if
$\ell$=3, and so on). We can then obtain the residual sequence
\begin{equation}
\epsilon_v(i)=u_v(i)-\widetilde{u}_v(i)~,~~~1\leqslant{i}\leqslant{s}~.
 \label{Eq:dfa:epsilon}
\end{equation}

{\em{Step 4}}. The detrended fluctuation function $F(v,s)$ of the
segment $u_v$ is defined as the root of the mean squares of the
sample residuals $\epsilon_v(i)$
\begin{equation}
 [F(v,s)]^2 = \frac{1}{s}\sum_{i = 1}^{s}[\epsilon_v(i)]^2~.
 \label{Eq:DFA_F}
\end{equation}
The overall detrended fluctuation is calculated by averaging over
all the segments, that is,
\begin{equation}
 [F(s)]^2 =\frac{1}{N_s}\sum_{v = 1}^{N_s}{[F(v,s)]^2}~.
 \label{Eq:DFA_Fs}
\end{equation}

{\em{Step 5}}. Varying $s$, we can determine the power-law relation
between the detrended fluctuation function ${F(s)}$ and the
timescale $s$,
\begin{equation}
F(s) \sim s^{H}~,
 \label{Eq:DFA:H}
\end{equation}
where $H$ is the DFA scaling exponent. In many cases including the
fractional Brownian motions, the DFA scaling exponent $H$ is
identical to the Hurst index
\cite{Taqqu-Teverovsky-Willinger-1995-Fractals,Kantelhardt-Bunde-Rego-Havlin-Bunde-2001-PA},
which is related to the power spectrum exponent $\eta$ by
$\eta=2H-1$ \cite{Talkner-Weber-2000-PRE,Heneghan-McDarby-2000-PRE}
and thus to the autocorrelation exponent $\gamma$ by $\gamma=2-2H$.

\subsection{The MFDFA algorithm}

The MFDFA method has the same first three steps as the DFA method,
and we need only to revise the last two steps
\cite{Kantelhardt-Zschiegner-Bunde-Havlin-Bunde-Stanley-2002-PA}.

{\em{Step 4}}. The $q$th order overall detrended fluctuation is
calculated as follows,
\begin{equation}
F_q(s) = \left\{\frac{1}{N_s}\sum_{v =
1}^{N_s}{[F(v,s)]^q}\right\}^{1/q}~,
 \label{Eq:DFA_M_Fqs}
\end{equation}
where $q$ can take any real value except for $q = 0$. When $q = 0$,
we have
\begin{equation}
F_0(s) = \exp\left\{\frac{1}{N_s}\sum_{v =
1}^{N_s}{\ln[F(v,s)]}\right\}~,
 \label{Eq:DFA_M_Fq0}
\end{equation}
according to L'H\^{o}spital's rule.

{\em{Step 5}}. Varying the value of $s$, we can determine the
power-law dependence of the detrended fluctuation function $F_q(s)$
on the size scale $s$, which reads
\begin{equation}
F_q(s) \sim s^{h(q)}~,
 \label{Eq:DFA_M_h}
\end{equation}
where $h(q)$ is the generalized Hurst index.

It is obvious that the DFA is a special case of the MFDFA when
$q=2$. In the standard multifractal formalism based on partition
function, the multifractal nature is characterized by a spectrum of
scaling exponents $\tau(q)$, which is a nonlinear function of $q$
\cite{Halsey-Jensen-Kadanoff-Procaccia-Shraiman-1986-PRA}. For each
$q$, we can obtain the corresponding traditional ${\tau(q)}$
function through
\begin{equation}
\tau(q) = qh(q) - D_f~, \label{Eq:MFDFA:tau}
\end{equation}
where $D_f$ is the fractal dimension of the geometric support of the
multifractal measure.

\subsection{The EMD algorithm}

Empirical mode decomposition is an innovative data processing
algorithm for nonlinear and non-stationary time series
\cite{Wu-Huang-Long-Peng-2007-PNAS}. It decomposes the time series
$x(t)$ into a finite number of intrinsic mode functions, which
satisfy the following two conditions: (1) in the whole set of data,
the numbers of local extrema and the numbers of zero crossings must
be equal or differ by 1 at most; and (2) at any time point, the mean
value of the ``upper envelope'' (defined by the local maxima) and
the ``lower envelope'' (defined by the local minima) must be zero.

The decomposing process is called a sifting process, which can be
described with the following six steps
\cite{Wu-Huang-Long-Peng-2007-PNAS}: (1) Identify all extrema of
$x(t)$; (2) Interpolate the local maxima to form an upper envelope
$U(x)$; (3) Interpolate the local minima to form a lower envelope
$L(x)$; (4) Calculate the mean envelope:
\begin{equation}
 \mu(t)=[U(x)+L(x)]/2;
 \label{Eq:EMD:mu}
\end{equation}
(5) Extract the mean from the signal
\begin{equation}
 g(t)=x(t)-\mu(t);
 \label{Eq:EMD:gt}
\end{equation}
and (6) Check whether $g(t)$ satisfies the IMF conditions. If YES,
$g(t)$ is an IMF, stop sifting; If NO, let $x(t)=g(t)$ and keep
sifting. Finally, we obtain
\begin{equation}
 r_n(t)=x(t)-\sum_{i=1}^n g_i(t),
 \label{Eq:EMD:rn}
\end{equation}
where $r_n$ is a residue representing the trend of the time series.

\subsection{The EMD-based DFA and MFDFA algorithms}

We can now embed the EMD algorithm into the DFA and MFDFA to modify
the third step of the algorithms, while keeping all other steps
unchanged.

{\em{Step 3}}. For each segment $u_v$, we obtain the EMD-based local
trend $\widetilde{u}_v=r_n(i)$ with the sifting process. We can then
obtain the residuals
\begin{equation}
\epsilon_v(i)=u_v(i)-r_n(i)~,~~~1\leqslant{i}\leqslant{s}~.
 \label{Eq:dfa:eps2}
\end{equation}
Note that, the trend $r_n(i)$ should be determined for each segment
separately at each timescale.

This gives the EMD-based DFA and MFDFA methods. One can see that
these methods differ essentially from the ones proposed in the
previous works \cite{Janosi-Muller-2005-PRE,Yeh-Fan-Shieh-2009-MEP}.

\section{Validating the methods through numerical experiments}
\label{s1:Simulation}

\subsection{EMD-based DFA of fractional Brownian motions}

We test the EMD-based DFA with synthetic fractional Brownian motions
(FBMs). In this paper, we use the free MATLAB software FracLab 2.03
developed by INRIA to synthesize fractional Brownian motions with
Hurst index $H_0$. In our test, we investigate fractional Brownian
motions with different Hurst indices $H$ ranging from 0.1 to 0.9
with an increment of 0.1. The size of each time series is
$2^{16}=65536$. For each $H_0$, we generate 100 FBM time series.
Each time series is analyzed by the EMD-based DFA algorithm.

\begin{figure}[htb]
\centering
\includegraphics[width=8cm]{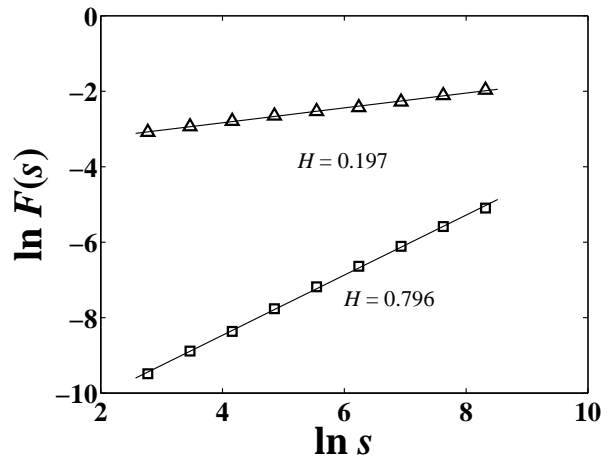}
\caption{\label{Fig:EMD:DFA:H28} Log-log plots of the detrended
fluctuation function $F(s)$ with respect to the timescale $s$ for
two randomly selected FBM time series with $H_0=0.2$ and $H_0=0.8$
using the EMD-based DFA algorithm. The solid lines are the
least-squares power-law fits to the data.}
\end{figure}

In Fig.~\ref{Fig:EMD:DFA:H28}, we show the log-log plot of the
detrended fluctuation $F(s)$ as a function of the timescale $s$ for
two randomly selected synthetic fractional Brownian motions with
$H_0=0.2$ (strongly anticorrelated) and $H_0=0.8$ (strongly
correlated), respectively. There is no doubt that the power-law
scaling between $F(s)$ and $s$ is very evident and sound, and the
scaling range spans more than two orders of magnitude. The estimates
of the Hurst indices are $H=0.197$ and $H=0.796$ for the
anticorrelated and correlated time series, which are very close to
the corresponding $H_0$ values. The EMD-based DFA algorithm is able
to well capture the self-similar (or self-affine) nature of the
fractional Brownian motions and results in precise estimation of the
Hurst index.

\begin{figure}[htb]
\centering
\includegraphics[width=8cm]{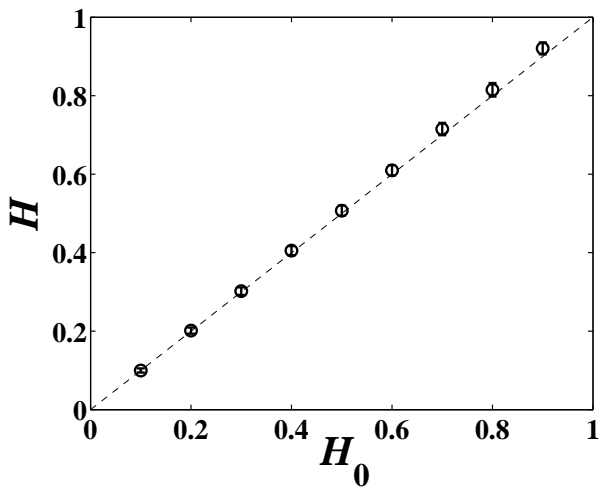}
\caption{\label{Fig:EMD:DFA:H:H0} Assessing the performance of the
EMD-based DFA method through extensive numerical experiments with
fractional Brownian motions. A comparison of the estimated Hurst
index $H$ with the true value $H_0$ is illustrated. The errorbar is
determined by the standard deviation of the 100 estimated $H$ values
for each $H_0$.}
\end{figure}

We confirm that there is also a power-law dependence of $F(s)$ on
$s$ for other synthetic FBMs with different Hurst index $H_0$. For
each $H_0$, we determine the Hurst index $H$ for each FBM time
series as done in Fig.~\ref{Fig:EMD:DFA:H28} and obtain 100 $H$
values. The mean of the 100 $H$ values is calculated for each $H_0$.
The resultant mean Hurst indices $H$ are plotted against $H_0$ in
Fig.~\ref{Fig:EMD:DFA:H:H0}. We can see that the estimated Hurst
indices $H$ are very close to the preset values $H_0$. The deviation
of the estimated Hurst index $H$ from $H_0$ becomes larger for
larger values of $H_0$. When comparing with the EMD variance method
which fails to give the estimates of Hurst index when $H_0<0.5$
\cite{Flandrin-Rilling-Goncalves-2004-IEEspl}, we find that the
EMD-based DFA method performs significantly better.

\begin{figure}[htb]
\centering
\includegraphics[width=8cm]{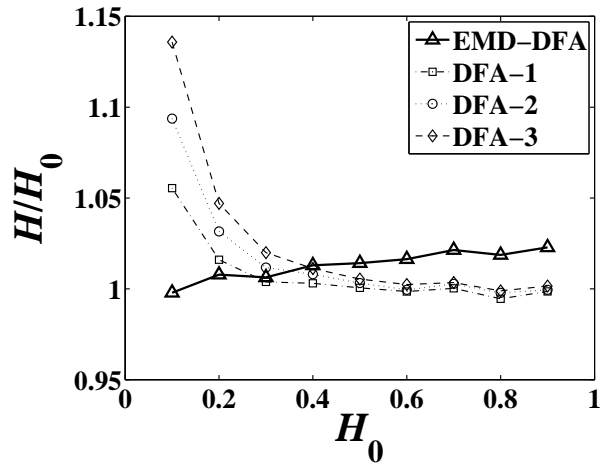}
\caption{\label{Fig:EMD:DFA:H:Perform} Assessing the performance of
the EMD-based DFA method with reference to the original DFA methods
using numerical experiments on fractional Brownian motions. The
relative ratio $H/H_0$ of the estimated Hurst index $H$ over the
true value $H_0$ is plotted as a function of $H_0$ for the EMD-based
DFA and the original DFA algorithms. }
\end{figure}

Although Fig.~\ref{Fig:EMD:DFA:H:H0} shows that the EMD-based DFA
method is able to determine the Hurst index of a given fractional
Brownian motion with high accuracy, it is necessary to compare the
performance of the EMD-based DFA algorithm with the original DFA
algorithm. We thus perform the DFA-1, DFA-2 and DFA-3 algorithms on
the same group of the FBM time series that were analyzed in
Fig.~\ref{Fig:EMD:DFA:H:H0}. The relative ratio of the estimated
Hurst index $H$ with reference to the true value $H_0$ is determined
for each $H_0$ for each of the four algorithms. Figure
\ref{Fig:EMD:DFA:H:Perform} digests the relative ratio $H/H_0$ as a
function of $H_0$ for the four algorithms. For the EMD-based DFA
algorithm, $H/H_0$ increases smoothly and its maximum is less than
1.03. It means that this method overestimates $H$ less than 3\%. In
contrast, the relative ratio $H/H_0$ of the DFA-$\ell$ algorithms
decreases with $H_0$ and increases with the polynomial order $\ell$.
For small $H_0$, the EMD-DFA method significantly outperforms the
DFA-$\ell$ methods. For DFA-3, the relative deviation is as large as
15\%. For large $H_0$, the DFA-$\ell$ methods do a better job than
the EMD-based DFA method.

\subsection{EMD-based MFDFA of multifractal signals}

We now turn to assess the performance of the EMD-based MFDFA
algorithm with synthetic multifractal signals. There are many
methods proposed to generate multifractal signals, such as the
multiplicative cascading method
\cite{Mandelbrot-1974-JFM,Meneveau-Sreenivasan-1987-PRL,Novikov-1990-PFA},
the fractionally integrated singular cascade method
\cite{Schertzer-Lovejoy-1987-JGR,Schertzer-Lovejoy-Schmitt-Chigirinskaya-Marsan-1997-Fractals,Decoster-Roux-Arneodo-2000-EPJB},
the random $\cal{W}$ cascades method
\cite{Arrault-Arneodo-Davis-Marshak-1997-PRL,Decoster-Roux-Arneodo-2000-EPJB},
and so on. The multiplicative cascading process is widely used to
model multifractal measures in many complex systems. As the simplest
one, the $p$ model was originally invented to simulate the
energy-dissipation field in turbulent flows
\cite{Meneveau-Sreenivasan-1987-PRL}. In this work, we adopt the
textbook $p$ model to generate multifractal signals.

We start from a line and partition it into two segments of the same
length and assign two given proportions of measure $p_1 = 0.3$ and
$p_2 = 1 - p_1$ to them. Then each segment is divided into two
smaller segments and the measure is redistributed in the same
multiplicative way. This procedure is repeated for 16 times and at
last we generate a multifractal signal of size $2^{16}=65536$. If
the multiplicative cascade process goes to infinity, the mass
exponent function has an analytic expression as follows
\begin{equation}
\tau(q) = -\ln(p_1^q + p_2^q)/\ln2~. \label{Eq:MFDFA:tau2}
\end{equation}
The empirical mass exponent function of the constructed multifractal
signal should be well approximated by Eq.~(\ref{Eq:MFDFA:tau2}).

\begin{figure}[htb]
\centering
\includegraphics[width=8cm]{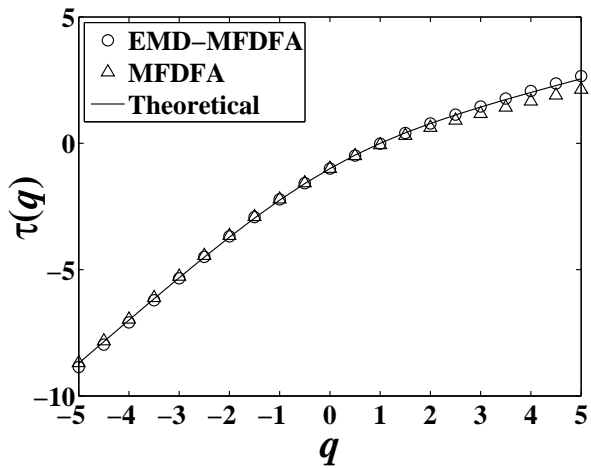}
\caption{\label{Fig:EMD:MFDFA:tau} Plots of $\tau(q)$ extracted from
the EMD-based MFDFA and the classical MFDFA as a function of $q$.
The continuous line is the theoretical formula
(\ref{Eq:MFDFA:tau2}).}
\end{figure}

We perform the EMD-based MFDFA on the binomial measure and determine
the empirical mass exponent function $\tau(q)$, which is illustrated
in Fig.~\ref{Fig:EMD:MFDFA:tau}. We also draw the theoretical line,
Eq.~(\ref{Eq:MFDFA:tau2}), in Fig.~\ref{Fig:EMD:MFDFA:tau} for
comparison. It is clear that the two curves overlap with each other.
In addition, we also perform the MFDFA on the same multifractal
signal and obtain the empirical $\tau(q)$ function, which is plotted
in Fig.~\ref{Fig:EMD:MFDFA:tau} as well. We observe a marked
discrepancy between the empirical $\tau(q)$ curve extracted based on
the classical MFDFA method and the theoretical curve when $q
\geqslant 2$. In addition, we find that the classical MFDFA method
systematically underestimates the $\tau(q)$ values when $q \geqslant
2$, which is also observed in the case of two-dimensional
multifractal surfaces \cite{Gu-Zhou-2006-PRE}. This test shows that
the EMD-based MFDFA method is able to extract the multifractal
nature of signals more accurate than the classical MFDFA method at
least in certain situations.

\section{Application to the SSEC index}
\label{S1:Application}

In this section, we apply the EMD-based MFDFA method to study the
multifractal nature of the high-frequency return time series of the
Shanghai Stock Exchange Composite (SSEC) index. The multifractal
properties of financial returns have been investigated extensively
\cite{Mantegna-Stanley-2000,Zhou-2007}. Concerning the Chinese stock
market, there are also numerous multifractal analyses based on  the
multiplier method \cite{Jiang-Zhou-2007-PA}, the MFDFA method
\cite{Du-Ning-2008-PA,Yuan-Zhuang-Jin-2009-PA}, and the partition
function approach
\cite{Wei-Huang-2005-PA,Du-Ning-2008-PA,Yuan-Zhuang-2008-PA,Wei-Wang-2008-PA,Jiang-Zhou-2008a-PA,Jiang-Zhou-2008b-PA}.
The presence of multifractality in the Chinese stock market is well
documented.

\begin{figure}[htb]
\centering
\includegraphics[width=8cm]{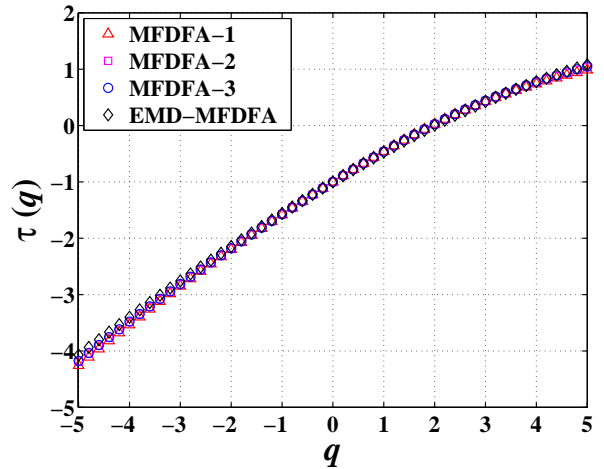}
\caption{\label{Fig:Tau:SSEC} (Color online) Multifractal analysis
of the 1-min high-frequency returns of the SSEC index from 4 January
2000 to 18 April 2008 using the EMD-based MFDFA and the classical
MFDFA methods.}
\end{figure}

We have performed the EMD-based MFDFA on the 1-min high-frequency
data of the SSEC index from 4 January 2000 to 18 April 2008. There
are 240 minutes in the double continuous auction on each trading
days \cite{Gu-Chen-Zhou-2007-EPJB,Gu-Chen-Zhou-2008a-PA}, and the
size of the data is 471202. For comparison, we have also conducted
the classical MFDFA with the polynomial order $\ell=1,2,3$. The
resulting $\tau(q)$ functions for $-5\leqslant {q}\leqslant 5$ are
illustrated in Fig.~\ref{Fig:Tau:SSEC}. It is evident that the four
algorithms give consistent results since the four $\tau(q)$ curves
overlap.

There are two intriguing characteristic points in
Fig.~\ref{Fig:Tau:SSEC}. When $q=2$, $\tau(q)=0$ for all the four
curves. It follows that the Hurst index of the SSEC returns is
$H=h(2)=[\tau(2)+1]/2=0.5$, which is in agreement with the
well-known fact that stock returns are uncorrelated. When $q=0$,
Fig.~\ref{Fig:Tau:SSEC} gives $\tau(q)=-1$, which is in line with
Eq.~(\ref{Eq:MFDFA:tau}). In certain sense, these two points verify
the correctness of the four algorithms.

\section{Summary}
\label{s1:Conclusion}

In summary, we have proposed a modified detrended fluctuation
analysis for fractal and multifractal signals based on the empirical
mode composition. The polynomial local trend in the classic DFA
algorithm is replaced by an EMD-based local trend. The modified
(MF)DFA is called the EMD-based (MF)DFA.

The performance of the EMD-based DFA and MFDFA methods is assessed
with extensive numerical experiments based on fractional Brownian
motion and multiplicative cascading process. For the EMD-based DFA
method, we investigated different Hurst index
$0.1\leqslant{H_0}\leqslant 0.9$ and generated 100 fractional
Brownian motions for each $H_0$. The accuracy of the estimated Hurst
indices $H$ obtained from the EMD-based DFA with reference to $H_0$
is compared with that from the classical DFA. We found that the
EMD-based DFA performs better than the classic DFA method in the
determination of the Hurst index when the time series is strongly
anticorrelated, especially when $H_0<0.3$, while the classical DFA
outperforms when $H_0>0.4$. In all cases, the EMD-based DFA is able
to determine the Hurst index $H$ with a deviation less than 3\% from
the true value $H_0$, that is $(H-H_0)/H_0<3\%$.

For the EMD-based MFDFA method, we constructed a multifractal signal
based on the $p$ model. The empirical mass exponent functions
$\tau(q)$ of the EMD-based MFDFA and classical MFDFA were compared
with the analytical expression. We found that the EMD-based MFDFA
outperforms the traditional MFDFA method when the moment order $q$
of the detrended fluctuations is positive. The usefulness of the
EMD-based MFDFA in the multifractal analysis is thus validated.

As an empirical example, we have applied the EMD-based MFDFA to the
1-min return data of Shanghai Stock Exchange Composite index. The
EMD-based MFDFA gives very similar results as the classical MFDFA
methods with different detrending polynomials. The presence of
multifractality is confirmed.

We conclude that the EMD-based DFA and MFDFA methods have comparable
performance as the classical DFA and MFDFA methods in the analysis
of fractal and multifractal time series. In certain cases, the
EMD-based methods can give better results. The only shortcoming of
the EMD-based DFA and MFDFA algorithms is that the detrending
process based on EMD is more time-consuming.

\begin{acknowledgements}
We are grateful to professor Zhaohua Wu for providing the Matlab
codes. This work was partially supported by the Program for New
Century Excellent Talents in University under grant NCET-07-0288 and
the Shanghai Educational Development Foundation under grant
2008SG29.
\end{acknowledgements}

\bibliography{E:/papers/Auxiliary/Bibliography}

\end{document}